


\input phyzzx
\hsize=6.5truein

\def\refmark#1{ [#1]}
\def\ack{\par\penalty-100\medskip \spacecheck\sectionminspace
   \line{\fourteenrm\hfil ACKNOWLEDGEMENTS\hfil}\nobreak\vskip\headskip }

\def\pr#1#2#3#4{{\it Phys. Rev. D\/}{\bf #1}, #2 (19#3#4)}
\def\np#1#2#3#4{{\it Nucl. Phys.} {\bf B#1}, #2 (19#3#4)}

\def\prl#1#2#3#4{{\it Phys. Rev. Lett.} {\bf #1}, #2 (19#3#4)}

\nopagenumbers
\line{\hfil CU-TP-673}
\line{\hfil gr-qc/9503035}
\vglue .5in
\centerline{\twelvebf Static Black Hole Solutions without Rotational
Symmetry}
\vskip .3in
\centerline{\it S.~Alexander Ridgway and Erick J.~Weinberg}
\vskip .1in
\centerline{Physics Department, Columbia University}
\centerline{New York, New York 10027}
\vskip .4in
\baselineskip=14pt
\overfullrule=0pt
\centerline {\bf Abstract}

We construct static black hole solutions that have no rotational
symmetry.  These arise in theories, including the standard electroweak
model, that include charged vector mesons with mass $m\ne 0$.  In such
theories, a magnetically charged Reissner-Nordstr\"om black hole with
horizon radius less than a critical value of the order of $m^{-1}$ is
classically unstable against the development of a nonzero vector meson
field just outside the horizon, indicating the existence of static
black hole solutions with vector meson hair.  For the case of unit
magnetic charge, spherically symmetric solutions of this type have
previously been studied.  For other values of the magnetic charge,
general arguments show that any new solution with hair cannot be
spherically symmetric.  In this paper we develop and apply a
perturbative scheme (which may have applicability in other contexts)
for constructing such solutions in the case where the
Reissner-Nordstr\"om solution is just barely unstable.  For a few low
values of the magnetic charge the black holes retain a rotational
symmetry about a single axis, but this axial symmetry disappears for
higher charges.  While the vector meson fields vanish exponentially
fast at distances greater than $O(m^{-1})$, the magnetic field and the
metric have higher multipole components that decrease only as powers
of the distance from the black hole.

\vskip .1in
\noindent\footnote{}{\twelvepoint \noindent This work was supported in part by
the US Department of Energy.}

\vfill\eject

\baselineskip=20pt
\pagenumbers
\pageno=1

\chapter{Introduction}

    One of the many remarkable features of black holes is the symmetry
and simplicity of the time-independent black hole solutions.  The
static vacuum black holes are all spherically symmetric and determined
by a single parameter.  Adding electromagnetism gives the possibility
of endowing the black hole with electric or magnetic charge, but the
static solutions remain spherically symmetric, with purely Coulomb
electromagnetic fields.  Even if one considers solutions that are
stationary, but not static, a rotational symmetry about one axis
remains.  This situation stands in constrast with that of
electromagnetism in flat spacetime, which possesses static (even if
singular) solutions
corresponding to point multipole moments of arbitrarily high order.
An explanation for the absence of static gravitational solutions with
higher multipoles comes from the no-hair theorems\ref{See, e.g.,
J.D.~Bekenstein, \pr5{1239}72.} that sharply constrain the possible
structure of black holes both in the electrovac case and for
gravity coupled to
a number of types of matter.

      However, it has become clear in recent years that if the theory
governing the matter fields has sufficient structure, it is in fact
possible to have black holes with nontrivial static fields outside the
horizon; i.e., black holes with hair.  In particular, theories with
electrically charged massive vector mesons can have two types of
magnetically charged black hole solutions\REFS\lnw{K.~Lee, V.P.~Nair
and E.J.~Weinberg, \pr{45}{2751}92.}\REFSCON\maison{P.~Breitenlohner,
P.~Forg\'acs, and D.~Maison, \np{383}{357}92.}\refsend.  One is the trivial
generalization of the Reissner-Nordstr\"om solution to the coupled
Einstein-Maxwell equations.  The other, which exists only if the
horizon radius is sufficiently small, has nonzero massive vector
fields just outside the horizon.  For the case of the $SU(2)$ gauge
theory with a triplet Higgs field, which has a nonsingular magnetic
monopole solution in flat spacetime, one finds a new solution with
unit magnetic charge that may be viewed as a Schwarzschild-like black
hole embedded in the center of an 't~Hooft-Polyakov
monopole\ref{G.~'t~Hooft, \np{79}{276}74; A.M.~Polyakov, {\it Pisma
v. Zh.  E.T.F.,} {\bf 20}, 430 (1974) [{\it JETP Lett.} {\bf 20}, 194
(1974)].}.  Although this solution was first found directly, a signal
of its existence is the fact that the Reissner-Nordstr\"om black hole
develops a classical instability when its horizon radius becomes
smaller than the radius of a magnetic monopole core\ref{K.~Lee,
V.P.~Nair and E.J.~Weinberg,
\prl{68}{1100}92.}.

      Similar arguments\Ref\kml{K.~Lee and E.J.~Weinberg,
\prl{73}{1203}94.} based on instabilities of Reissner-Nordstr\"om
solutions suggest the existence of new black hole solutions with
higher magnetic charges.  However, in the presence of a magnetic
monopole a spherically symmetric charged spin-one field is
possible\REFS\guth{A.H.~Guth and E.J.~Weinberg,
\pr{14}{1660}76.}\REFSCON\harmonics{E.J.~Weinberg,
\pr{49}{1086}94.}\refsend\
only if the product of the magnetic charge of the monopole and the
electric charge of the field is unity\rlap.\footnote{1}{This follows
from the absence of vector spherical harmonics with total angular
momentum zero.  The absence of such harmonics can be understood by
considering the motion of a particle with electric charge $e$ about a
monopole with magnetic charge $Q_M$.  The total angular momentum is
the sum of the spin angular momentum, the orbital angular momentum,
and a contribution of magnitude $eQ_M$ directed along the line from
the monopole to the charge; if $eQ_M \ne 1$, the sum of these three
terms can never vanish.}  Hence, these new black holes can be at most
axially symmetric.  Whether or not they actually possess such symmetry
is a question not of general principle, but of detailed dynamics.  In
this paper we will show that, at least for certain ranges of
parameters, they do not.

      A theory with sufficient structure to yield these new black
holes has matter fields described by the flat-spacetime Lagrangian
$$  \eqalign{ {\cal L} &= -{1\over 4}F_{\mu\nu}F^{\mu\nu}
        -{1\over 2}  W^*_{\mu\nu}W^{\mu\nu}
        -  m^2 W^*_\mu W^\mu \cr   &\qquad
     -{ieg\over 4} F^{\mu\nu} \left( W^*_\mu W_\nu -W^*_\nu W_\mu\right)
     -{\lambda e^2 \over 4} \left| W^*_\mu W_\nu -W^*_\nu W_\mu\right|^2}
   \eqn\lagrangian $$
where
$$   F_{\mu\nu} = \partial_\mu A_\nu - \partial_\nu A_\mu
\eqno\eq  $$
$$   W_{\mu\nu} = D_\mu W_\nu - D_\nu W_\mu
\eqno\eq  $$
$$ D_\mu W_\nu = (\partial_\mu -ie A_\mu) W_\nu \, .
   \eqno\eq $$
The fourth term in the Lagrangian is an anomalous magnetic moment
term, with the constant $g$ arbitrary.  In order that the energy be
bounded from below, we must require that $\lambda \ge g^2/4$\refmark\kml.

     If we add to the theory a neutral scalar field $\phi$ with
appropriate self-interactions and give the vector field a
$\phi$-dependent mass $m=e\phi$, then for $g=2$ and $\lambda=1$, then
Eq.~\lagrangian\ is simply the unitary gauge form of the Lagrangian for
an $SU(2)$ gauge theory spontaneously broken to $U(1)$ by a triple
Higgs field.  Similarly, for $g=2$, $\lambda = 1/\sin^2\theta_W$, and
$m=e\phi/2$ we obtain the unitary gauge form of the standard
electroweak Lagrangian, but with all terms involving the $Z$ or
fermions omitted.  It is a straightforward matter to extend the
analysis of this paper to such models.

     We are seeking static black hole solutions to the theory obtained
by coupling the Lagrangian of Eq.~\lagrangian\ to general relativity.
In the absence of rotational symmetry, the static field equations are
a set of coupled partial differential equations in three variables.
An exact analytic solution of these is beyond our abilities.  Instead,
we use a perturbative approach.  We begin by considering a
Reissner-Nordstr\"om black hole with radial magnetic field
$$  F_{\theta\phi} = {q\over e} \sin\theta
    \eqn\RNmagfield $$
corresponding to a magnetic charge $Q_M =q/e$ (where $q$ is restricted
by the Dirac quantization condition to integer or half-integer values)
and vanishing $W$ field.  The metric is
$$ ds^2 = - B(r) dt^2 + B^{-1}(r) dr^2
    + r^2 (d\theta^2 + \sin^2\theta d\phi^2)
    \eqno\eq $$
with
$$  B(r)  = 1 - {2MG \over r} + {4\pi G q^2 \over e^2r^2}
   = {(r-r_H)(r-r_-) \over r^2}  \, .
    \eqno\eq $$

   We choose the mass $M$ to be such that the outer horizon radius $r_H$
is less than $r_{\rm cr}$, the critical value for instability.  It is in
this mass range that we expect there to be a second black hole solution
with nontrivial $W$ field and with electromagnetic field strengths and
metric that differ from the Reissner-Nordstr\"om form.  It is often the
case that the exponentially growing eigenmodes about an unstable static
solution give a good indication of the nature of a nearby stable
solution, particularly in the case where the original solution is just
barely unstable.   Guided by this intuition, we linearize the static
field equations about the Reissner-Nordstr\"om solution.  This leads to an
eigenvalue problem that is closely related to, although not exactly the
same as, that encountered in the stability
analysis\Ref\Alex{S.A.~Ridgway and E.J.~Weinberg, \pr{51}{638}95.}.
For $r_H$ close to
$r_{\rm cr}$, there is a single negative eigenvalue, whose magnitude
tends to zero as $r_H \rightarrow r_{\rm cr}$.   With $r_{\rm cr} - r_H$
sufficiently small, this eigenvalue becomes a small parameter that can
serve as the basis for a perturbative expansion.

In Sec.~2, we illustrate our method with a simple toy model consisting
of a scalar field coupled to a fixed, but spatially inhomogneous,
external source.  In Sec.~3, we set up the formalism for treating the
case in which are actually interested, that of a black hole in the
theory described by the Lagrangian of Eq.~\lagrangian.  We assume that
$Gm^2/e^2$ is small; this allows one to solve (to leading order) for the
charged vector field and the perturbations of the electromagnetic
field before dealing with the metric perturbations.  For technical
reasons, it turns out that the details of the subsequent analysis are
considerably simpler if $g$ is positive and $q \ge 1$.  We exploit
these simplifications in Sec.~4, where we determine the leading
perturbations of the electromagnetic field in terms of those of
$W_\mu$.  In Sec.~5, we examine the lowest order contributions to the
charged vector field and show that there is a parameter range for
which the solution is not even axially symmetric.  In Sec.~6 we obtain
the leading corrections to the Reissner-Nordstr\"om metric.  Some
concluding remarks are included in Sec.~7.  An appendix describes some
results needed to construct Green's functions that we use.

\chapter{A Toy Model}

    Consider a real scalar field whose dynamics is governed by
the Lagrangian
$$ {\cal L} = -{1\over 2} \left(\partial_\mu \phi \right)^2
    - {1\over 2} F({\bf x}) \phi^2 -{\lambda\over 4} \phi^4
    \eqno\eq $$
where $ F({\bf x})$ arises from the coupling to a spatially inhomogeneous
but static external source.   Static solutions obey
$$ \eqalign{  0 &= [- \nabla^2 + F({\bf x})]\phi + \lambda \phi^3 \cr
     &\equiv {\cal M} \phi + \lambda \phi^3 }
     \eqn\static $$
and, to have finite energy, must satisfy the boundary condition
that $\phi$ vanish at spatial infinity.

     The trivial configuration $\phi(x)=0$ is a solution for
any choice of $F({\bf x})$.
It is easy to see that this is the only
static solution if $\cal M$ is a positive operator.  However, if
$\cal M$ has negative eigenvalues, this trivial solution is
unstable, implying the existence of a new, spatially
inhomogeneous, solution whose form we seek.  To this end, let us
assume that $\cal M$ has only a single negative eigenvalue, with an
eigenfunction $\psi$ obeying
$$  {\cal M} \psi = -b^2 \psi
    \eqno\eq $$
and normalized so that
$$  \int d\,^3x\, \psi^2({\bf x})  = 1 \, .
    \eqno\eq $$

    We now write $\phi$ as the sum of a term proportional to $\psi$ and
one orthogonal to it; i.e.,
$$ \phi(x) = k \psi(x) + \tilde \phi(x)
    \eqno\eq $$
with
$$  \int d\,^3x\, \psi({\bf x}) \tilde \phi({\bf x}) = 0 \, .
    \eqn\orthog $$
The static field equation \static\ then implies that
$$  {\cal M} \tilde \phi + \lambda \left( k \psi +\tilde \phi\right)^3
     +\Gamma \psi =0
    \eqn\tilphi $$
where $\Gamma$ is a Legendre multiplier that enforces the
orthogonality condition \orthog.  It can be calculated by multiplying
both sides of this
equation by $\psi$ and then integrating over all space to obtain
$$  \Gamma = -\lambda
          \int d\,^3x \,\psi \left( k \psi +\tilde \phi\right)^3 \, .
    \eqn\phigamma $$
Variation of the action with respect to $k$
gives the additional equation
$$   {\partial I \over \partial k} =0
       \eqn\phik $$
where
$$  \eqalign{ I &= \int d\,^3x \left[ {1\over 2} k^2 \psi{\cal M}\psi
    + {\lambda \over 4} \left( k \psi +\tilde \phi\right)^4 \right] \cr
        &= - {1\over 2} k^2 b^2 + {\lambda \over 4} \int d\,^3x
        \left( k \psi +\tilde \phi\right)^4 \, .}
       \eqn\Itoy $$

    Thus far we have made no approximations; together, Eqs.~\tilphi,
\phigamma, and \phik\ are completely equivalent to Eq.~\static.  We now
recall that if $b=0$ the scalar field $\phi$, and therefore $k$ and
$\tilde \phi$, must vanish.  Hence, for small $b$ it
should be possible to expand these quantities as power series in
$b$.  Furthermore, since it is the existence of the negative
eigenvalue which makes a nontrivial solution possible, we may view $k\psi$
as providing the source for $\tilde\phi$ (through Eq.~\tilphi).  We
therefore expect
that $\tilde \phi$ is of higher order in $b$ than $k
\psi$.  Assuming this to be the case, Eqs.~\phik\ and \Itoy\ give
$$ k^2 = {b^2 \over \lambda}
     \left[ \int d\,^3x \psi^4({\bf x}) \right]^{-1}  + O(b^3)
     \eqno\eq  $$
while Eq.~\phigamma\ implies
$$ \eqalign{ \Gamma & = -\lambda  k^3 \int d\,^3x \psi^4({\bf x})
           +O(b^4) \cr
    &= {b ^3 \over \sqrt{\lambda} }
  \left[ \int d\,^3x \psi^4({\bf x}) \right]^{-1/2}  + O(b^4) \, . }
     \eqno\eq  $$
These may be substituted into Eq.~\tilphi\ to give
$$ \eqalign{ {\cal M} \tilde \phi({\bf x}) &= -\lambda k^3
   \left[ \psi^3({\bf x}) - \psi({\bf x}) \int d\,^3y \psi^4({\bf y})
    \right]  + O(b^4) \cr
   &= {b ^3 \over \sqrt{\lambda} }
      \left[ \int d\,^3x \psi^4({\bf x}) \right]^{-3/2}
 \left[ \psi^3({\bf x}) - \psi({\bf x}) \int d\,^3y \psi^4({\bf y})
    \right]  + O(b^4) \, .}
  \eqno \eq $$
This shows that $\tilde \phi$ is of order $b^3$, and justifies the
assumption above that it is of higher order than $k\psi$.  Thus, to
leading order the static solution is approximated by the negative
eigenvalue fluctuation about the vacuum solution, multiplied by a
scale factor whose magnitude is determined by the nonlinear term in
the Lagrangian.

By substitution of the lower order results back into the original
equations, $k$ and $\tilde\phi$, and hence $\phi$ itself, can be
calculated to to arbitrarily high order in $b$.

\chapter{Charged Vector Meson Model}

  We now apply this method to the theory in which we are actually
interested, namely that described by the Lagrangian of
Eq.~\lagrangian.  The first step is to identify the unstable (i.e.,
exponentially growing in time) modes about the unperturbed
Reissner-Nordstr\"om solution.  This was done in Ref.~\Alex, whose results
we now briefly summarize.
When the field equations are linearized, the perturbations
in the gauge field and the metric decouple from those in the massive
vector field.  The linear perturbation problem for the former two
modes is the same as in the pure Einstein-Maxwell theory, where it was
shown some time ago\ref{V.~Moncrief, \pr{9}{2707}74; \pr{10}{1057}74;
\pr{12}{1526}75.} that the Reissner-Nordstr\"om solution is stable.
Hence, the stability analysis reduces to a study of the
linearized $W$ field equations.  It is convenient to define
${\cal M}_{\mu\nu}$ by
$$ {\cal M}^{\mu\nu} W_\nu =
    - {1\over \sqrt{\bar g}} \bar D_\alpha \left(\sqrt{\bar g}
W^{\alpha\mu}\right)
     +m^2 W^\mu  - {ieg \over 2}\bar F^{\alpha\mu}W_\alpha
    \eqno\eq $$
where here, and for the remainder of the paper, we adopt
the convention that $\bar g_{\mu\nu}$,
$\bar A_\mu$, and $\bar F_{\mu\nu}$ denote the corresponding
unperturbed quantities while $\bar D_\mu$ is the gauge
covariant derivative taken with respect to the unperturbed potential;
indices are raised and lowered with the unperturbed metric.  The
unstable modes are solutions of
$$ {\cal M}_\mu{}^\nu W_\nu = 0
    \eqn\stability $$
whose time-dependence is of the form
$$  W_\mu({\bf x},t) = f_\nu({\bf x}) e^{\omega t}
   \eqn\timedepmode $$
with real $\omega$.  The spherical symmetry of the unperturbed solution
allows one to choose  the solutions of Eq.~\stability\ to be
eigenfunctions of both ${\bf J}^2$ and $J_z$, where $\bf J$ is the total
angular momentum operator.  Because of the extra angular momentum of
an electric
charge in the  field of a magnetic monopole, the corresponding
eigenvalues are not the usual ones.  Instead, $J$ runs in integer steps
upward from the minimum value $J_{\rm min}= q-1$, unless $q=1/2$, in
which case $J_{\rm min}=1/2$. For each value of $J$, unstable modes
exist if the horizon radius $r_H$ is less than a critical value
$r_{\rm cr}(J)$ that is of order $m^{-1}$, provided
that $g$ lies in an appropriate range ($g>0$ for $J=q-1$, $g>2$ for
$J=q$, and either $g<0$ or $g>2$ for $J>q$).  For a given value of
$g$, $r_{\rm cr}(J)$ is greatest for the smallest $J$ that can
have unstable modes with that $g$.  Thus, if $\hat J$ denotes the value
that maximizes $r_{\rm cr}$, we have $\hat J = q-1$ for $g>0$ and $\hat
J= q+1$ for $g<0$ for $q \ge 1$. If $q=1/2$, $\hat J = 1/2$ if $g>2$ and
$3/2$ if $g<0$; if $0\le g \le 2$, there is no instability.

      The modes that will form the basis for our new solutions
are the static eigenfunctions of $\cal M$
with negative eigenvalue; i.e., the time-independent solutions of
$$ {\cal M}_\mu{}^\nu \psi_\nu = -\beta^2 m^2 \psi_\mu
    \eqn\eigenvalue $$
with real $\beta$.  (A factor of $m^2$ has been extracted to make
$\beta$ dimensionless.)

   This eigenvalue equation must be supplemented by boundary conditions.
At spatial infinity we merely require that $\psi_\mu$ not diverge.
For negative eigenvalues (indeed for all eigenvalues less than $m^2$)
this implies that $\psi_\mu$ in fact vanishes as $r\rightarrow
\infty$.  A second boundary condition is obtained at the horizon, where
we require that $\psi_\mu$ be regular, in the sense that its
components measured relative to a coordinate
system that is nonsingular at the horizon (e.g., Kruskal-like
coordinates) be regular.   Because of the manner in which the
singular metric factors enter Eq.~\eigenvalue, this
constrains the behavior of $\psi_\mu$ near the horizon --- as we will
see more explicitly in the next section --- and causes the
spectrum of negative eigenvalues to be discrete\rlap.\footnote{2}{It
might seem
strange that the nature of the spectrum should be determined by the
singularities of a metric at a horizon that is only a coordinate
singularity.  This happens because the condition we are imposing on
the eigenfunctions, that they be static, is defined in terms of a
coordinate $t$ that is singular at the horizon.}   For this
portion of the spectrum, we can require that the eigenfunctions
satisfy the normalization condition\footnote{3}{Since, as is easily
shown, static solutions of Eq.~\eigenvalue\ must have $\psi_t=0$,
$\psi_\mu\psi^\mu$ is positive.}
$$  \int d\,^3x\sqrt{\bar g}\, \psi^*_\mu\psi^\mu   = 1
    \eqn\psinorm $$
where, both in this equation and hereafter, the spatial integration
is understood to be restricted to the region outside the
Reissner-Nordstr\"om outer horizon.

     Because of both the nontrival metric component
$g_{tt}(r)$ and the possibility of a nonvanishing $W_t$ in the
time-dependent case, the eigenfunctions $\psi_\mu$ are not in general
the same as the $f_\mu$ that appear in Eq.~\timedepmode; the
spectra of the $\omega$ and $\beta$ are not even the same.   However,
a zero eigenvalue for the static operator does correspond to a zero
frequency of the small oscillation problem and, furthermore, the static
problem has negative eigenvalues if and only if the Reissner-Nordstr\"om
solution is unstable.   Hence, the conditions for instability
enumerated above are also the conditions we need to be able to construct
our new solutions.

    In fact, the static eigenmodes for real $\beta$ can be obtained
from the $\omega=0$ solutions of Eq.~\stability\ with different values
for the parameters.   This can be seen by bringing the right hand side
of Eq.~\eigenvalue\ over to the left; the resulting equation is precisely that
satisfied by $f_\mu({\bf x})$ for $\omega=0$, but with $m^2$ replaced
by $m^2(1+\beta^2)$.   It follows that the value of $r_H$ that leads to
a given $\beta$ for $W$-mass $m$ is equal to the critical value $r_{\rm
cr}$ for a $W$-mass $m\sqrt{1+\beta^2}$.  If $m \ll M_{\rm Pl}$ (the
case with which we will be primarily concerned), $r_{\rm cr}$ is much
greater than the horizon size for an extremal Reissner-Nordstr\"om black
hole and its dependence on the inner horizon $r_-$ can be neglected.
Dimensional arguments then show that $r_{\rm cr}$ is inversely
proportional to $m$.  It then follows that
$$  \beta = {\sqrt{r_{\rm cr}^2 - r_H^2} \over r_H} \, .
     \eqno\eq $$
Thus, by taking  $r_H - r_{\rm cr}(\hat J) \ll
r_{\rm cr}(\hat J)  $, we ensure that $\beta \ll 1$, thus providing
the small parameter needed for our perturbative calculation.

   In general $\hat J$ is nonzero, so that
instead of a single unstable mode, as in the model of Sec.~2, there is
a degenerate multiplet of unstable modes $\psi^M_\mu$ that are
distinguished by the eigenvalue of $J_z$.  Proceeding as in that
section, we
write the $W$ field as a linear combination of the unstable
modes plus a remainder orthogonal to these modes,
$$ W_\mu  = V_\mu + \tilde W_\mu
   = m^{-1/2} \sum_{M= -\hat J}^{\hat J} k_M \psi^M_\mu + \tilde W_\mu
\, ,
   \eqn\Wexpand $$
where
$$   \int d\,^3x\sqrt{\bar g}\, \psi^*_\mu\tilde W^\mu   = 0 \, .
    \eqno\eq $$

It is useful to define a quantity $a$ by
$$  \sum_{M=-\hat J}^{\hat J} |k_M|^2 =  a^2
    \eqn\adef$$
so that
$$ V_\mu = O(a) \, .
     \eqno\eq$$
Since, will be displayed explicitly below, the source for the
perturbations of the electromagnetic field is quadratic in $V_\mu$ and
contains an explicit factor of $e$,
$$ \delta A_\mu = O(ea^2) \, .
    \eqno\eq $$
(Note that the quantities $ea$ and
$Gm^2/e^2$ are truly dimensionless, whereas $e$ is dimensionless only
if one sets $\hbar=1$, which would not be natural in this essentially
classical context.)
The source for the metric perturbations is the perturbation of the
energy-momentum tensor.  The leading contribution to this, of order
$a^2$, is from terms quadratic in $V_\mu$ and from terms linear in
$\delta A_\mu$.  However, these enter the field equation multiplied by
a factor of $G$, and so $\delta g_{\mu\nu} \equiv h_{\mu\nu}$ must be
suppressed by an additional factor of roughly $Gm^2 = (m/M_{\rm
Pl})^2$. Hence,
$$  \delta g_{\mu\nu} \equiv h_{\mu\nu} = O(Gm^2 a^2) \, .
    \eqno\eq $$

Finally, the magnitude of $\tilde W$ can be determined from the field
equation
$$ \eqalign{ {\cal M}^{\nu\alpha} \tilde W_\alpha &=
   - \lambda e^2 \left(V^{*\mu}V^\nu -  V^{*\nu}V^\mu \right) V_\mu
     +{ieg\over 2} V_\mu \delta F^{\mu\nu}
       -ie \delta A_\mu \left(\bar D^\mu V^\nu -\bar D^\nu V^\mu \right) \cr
  &\quad      +{ie\over \sqrt{\bar g}}  \bar D_\mu\left[ \sqrt{\bar g}
  \left( V^\mu \delta A^\nu -  V^\nu \delta A^\mu \right)\right]
    - \sum \Gamma_M \psi_M^\nu   + \cdots \, . }
        \eqn\Weq $$
Here the dots represent terms which are either $O(e^4a^5)$ or $O(Gm^2 a^3)$ or
smaller, while the
$\Gamma_M$ are Lagrange multipliers introduced to enforce the
orthogonality of  $\tilde W_\mu$ and the $\psi_\mu^M$.  One can solve
for the $\Gamma_M$ by multiplying
both sides of this equation by $\psi^M_\mu$ and then integrating over
all space outside the horizon.  Inserting the result back into Eq.~\Weq,
we see that
$$ \tilde W_\mu = O(e^2a^3) \, .
   \eqno\eq $$

    We will assume that $Gm^2/e^2 \ll 1$.
The leading behavior of $V_\mu$, $\tilde W_\mu$, and $\delta A_\mu$
can then be obtaining by solving the field equations in the background
of the unperturbed Reissner-Nordstr\"om metric.  Having done this, the
leading perturbations of the metric can then be obtained.  In fact,
for calculating the lowest order metric perturbations,
only $\delta A_\mu$ and $V_\mu$ are needed.  For the remainder of this
section, and the next two, we will concentrate on the determination of
these two quantities.  We will then return to the calculation of the
metric perturbations in Sec.~6.

    Linearization of the electromagnetic field equation about the
unperturbed solution yields
$$ {1\over \sqrt{\bar g}} \partial_\mu \left(\sqrt{\bar g}
   \, \delta F^{\mu\nu} \right) =
    {1\over \sqrt{\bar g}} \partial_\mu \left(\sqrt{\bar g} \,
         p^{\mu\nu} \right) +j^\nu
   \eqn\Feq $$
where
$$  p_{\mu\nu} = -{ieg\over 2} \left(V^*_\mu V_\nu -V^*_\nu V_\mu \right)
       + \cdots
  \eqno\eq $$
and
$$ j^\nu = ie \left[V_\mu^* \left(\bar D^\mu V^\nu -\bar D^\nu V^\mu \right)
      - V_\mu \left(\bar D^\mu V^{*\nu} -\bar D^\nu V^{*\mu} \right) \right]
       +\cdots
   \eqno\eq $$
with the dots signifying higher order terms.
Similarly, the Bianchi identity gives
$$ \epsilon^{\mu\nu\alpha\beta}  \partial_\nu \delta F_{\alpha\beta}
=0 \, .
     \eqn\EMbianchi $$

    In addition to these, we need the equations obtained by
varying the action with respect to the $k_M$.   In deriving these, we
need only take
into account terms in the action of up to order $e^2a^4$ and so can
approximate the action by
$$ \eqalign{ S_{approx} &= \int d\,^4x \sqrt{\bar g}
    \left[ - V^*_\mu {\cal M}^{\mu\nu} V_\nu -
      {\lambda e^2\over 4} \left| V^*_\mu V_\nu -V^*_\nu V_\mu\right|^2
           \right. \cr &\qquad \left.
      -{1\over 4} \left(\bar  F_{\mu\nu} + \delta F_{\mu\nu} \right)
         \left(\bar  F^{\mu\nu} + \delta F^{\mu\nu} \right)
      +{1\over 2} \delta F_{\mu\nu} p^{\mu\nu}  - \delta A_\nu j^\nu
       \right] \, .}
    \eqn\Sapprox $$
(Terms linear in both $\tilde W_\mu$ and $V_\nu$, which would be of
order $e^2a^4$, are absent because of the orthogonality of $\tilde
W_\mu$ and the $\psi^M_\mu$.) There are two ways of proceeding from
here.  One can substitute the expansion of Eq.~\Wexpand\ for $V_\mu$
and then vary the above expression with respect to $k_M$, thus
obtaining an equation involving both the $k_M$ and $\delta A_\nu$.
Alternatively, one can first use Eqs.~\Feq\ and \EMbianchi\ to solve
for $\delta A_\mu$ and $\delta F_{\mu\nu}$ in terms of $V_\mu$ and
then substitute the resulting expressions back into Eq.~\Sapprox\ to
obtain an action which is a function of only the $k_M$; we will follow
this second approach.  A number of simplifications are possible.
First, the eigenvalue equation \eigenvalue\ and the normalization
condition \psinorm\ can be used to integrate the term quadratic in
$V_\mu$.  Next, by multiplying both sides of Eq.~\Feq\ by $\sqrt{\bar
g}\, \delta A_\nu$ and then integrating by parts, one obtains the identity
$$   \int d\,^4x \sqrt{\bar g} \, \delta F_{\mu\nu}\,\delta F^{\mu\nu}  =
  \int d\,^4x \sqrt{\bar g} \left[ \delta F_{\mu\nu} p^{\mu\nu}
         - \delta A_\nu j^\nu \right]
     \eqno\eq $$
which can be used to eliminate the term quadratic in $\delta F_{\mu\nu}$.  A
similar procedure applied to the source-free equation obeyed by the
unperturbed field strength shows that
$$   \int d\,^4x \sqrt{\bar g} \, \bar F_{\mu\nu}\, \delta F^{\mu\nu}
= 0 \, .
     \eqno\eq $$
(In both cases, one can verify that the surface terms from the integration by
parts vanish as long as the total magnetic charge is held fixed.)
The term quadratic in $F_{\mu\nu}$ is obviously independent of the
$k_M$ and can be ignored.   Finally, since all quantities are independent of
time, it is sufficient to integrate over the spatial variables.  We
are thus led to the equation
$$   0 = {\partial I \over \partial k_M}
    \eqno \eq $$
where
$$  I =  -\beta^2 m a^2
    +  \int d\,^3x \sqrt{\bar g} \left[
       {\lambda e^2\over 4} \left| V^*_\mu V_\nu -V^*_\nu V_\mu\right|^2
      -{1\over 4} \delta F_{\mu\nu} p^{\mu\nu}
    + {1\over 2}\delta A_\nu  j^\nu \right]
    \eqn\Igeneraleq $$
and $\delta A_\nu$ is understood to be given in terms of the $k_M$.
Since the integrand on the right hand side is of order $e^2a^4$, we see
that $a$ is proportional to $\beta/e$, indicating that our perturbative
expansion is justified for $r_H$ sufficiently close to $r_{\rm cr}$.

\chapter{The Case $q \ge 1$, $g>0$}

     We now specialize to the case $q \ge 1$, $g>0$ for which, as was
noted above, $\hat J = q-1$.  This allows us to take advantage of the
special properties\refmark\harmonics\ of the $J=q-1$ vector spherical
harmonics, which
lead to a number of technical simplifications in the analysis.
\REF\olsen{H.A.~Olsen, P.~Osland, and T.T.~Wu, \pr{42}{665}90.}
For $J=q-1$, and only for that value, there is but a single
monopole vector spherical harmonic\refmark{\harmonics,\olsen} for
each value of $J_z =M$.  Hence, if we denote this harmonic by
$C_\mu^M(\theta,\phi)$, the unstable modes in this case can be written
in the form
$$  \psi_\mu^M   = f(r)\,  C_\mu^M(\theta,\phi)
    \eqno\eq $$
where $f(r)$ does not depend on $M$.

    The $J=q-1$ harmonics have a number of special properties.  Their
radial and time components vanish,
$$  C_r^M = C_t^M =0 \, ,
    \eqn\Crt $$
and  their two angular components are related by
$$  C_\phi^M = i \sin\theta \, C_\theta^M \, .
    \eqn\Cangular  $$
In addition, their covariant curl, evaluated in the
background Dirac vector potential, vanishes:
$$ \bar D_\mu C_\nu^M - \bar D_\nu C_\mu^M =0
    \eqno\eq  $$
as does their covariant divergence
$$ {1\over \sqrt{\bar g} }\bar D_\mu \left( \sqrt{\bar g} C^{M\mu}
     \right)  =0 \, .
     \eqno\eq $$
A convenient choice of normalization condition is
$$  \int d\phi\, d\theta \sin\theta
     \left[ C^{M}_\mu(\theta,\phi)\right]^*  C^{M \mu}(\theta,\phi)
            =  {1\over r^2}  \, .
     \eqn\Cnorm $$

     To obtain an explicit expression for the $C^{M}_\mu$, we must
choose a gauge.  If the electromagnetic vector potential
has a single nonvanishing component
$$ A_\phi = {q\over e} (1-\cos\theta) \, ,
    \eqno\eq $$
then
$$    C_\theta^M =  a_{qM}e^{i\phi} (1+ \cos \theta)^{q-1}
    \left[ {\sin\theta \over 1+\cos\theta} e^{i\phi} \right]^{q+M-1}
     \eqn\explicit $$
where
$$  a_{qM} = {1\over 2^q \sqrt{2\pi}}
   \left[{(2q-1)! \over (q+M-1)! (q-M-1)!} \right]^{1/2} \, .
    \eqno\eq $$

    With the aid of these properties, the eigenvalue equation
\eigenvalue\ reduces to
$$   -{d\over dr} \left(B {df\over dr} \right)
   +\left( m^2 -{qg \over 2r^2} \right) f     = -\beta^2 m^2 f
     \eqn\feq $$
where $f(r)$ can be chosen to be real.  Equations~\psinorm\ and
\Cnorm\ fix the normalization of $f$ to be
$$  \int_{r_H}^\infty dr   |f(r)|^2 = 1  \, .
     \eqn\fnorm $$
Given $r_{\rm cr}$, and hence $\beta$, Eq.~\feq\ can be integrated
numerically to obtain $f(r)$.  This function is monotonic, has no
zeros, and  vanishes exponentially with $r$ as
$r\rightarrow\infty$.  Near the horizon it behaves as
$$ f(r)  = A [ 1 - b(r-r_H) ] + O[(r-r_H)^2]
     \eqno \eq $$
where
$$   b = \left[{qg\over r_H^2} -m^2(1+\beta^2) \right]
        \left[ B'(r_H) \right]^{-1}  >0 \, .
     \eqno\eq $$

  Proceeding with the construction of the solution, we write
$$  V_\mu = m^{-1/2} f(r) \Phi_\mu (\theta,\phi)
    \eqno\eq $$
where
$$  \Phi_\mu(\theta,\phi) = \sum_{M=-(q-1)}^{q-1}
         k_M C_\mu^M(\theta,\phi)  \, .
    \eqn\Phidef $$
Eqs.~\Crt\ and \Cangular\ imply that $\Phi_r=\Phi_t=0$ and fix the
ratio of $\Phi_\theta$ and $\Phi_\phi$.   Note that
$\Phi_\mu$ has exactly $2(q-1)$ zeros as $\theta$ and $\phi$ range over
the unit sphere.  To show this, we use the explicit expression
\explicit\ for the vector harmonics and write
$$  \Phi_\theta (\theta,\phi) = e^{i\phi} (1+\cos\theta)^{q-1}
      \sum_{M=-(q-1)}^{q-1} a_{qM} k_M z^{q+M-1}
   \eqn\Phisum $$
where
$$   z= {\sin\theta \over 1+\cos\theta} e^{i\phi} \, .
    \eqno\eq  $$
The entire complex $z$-plane maps onto the unit sphere, with $|z|
=\infty$ corresponding to the south pole, $\theta =\pi$.  Let $\bar M$
be the largest value of $M$ for which $k_M$ is nonzero.  The sum in
Eq.~\Phisum\ is then a polynomial of order $\bar M +q-1$ in $z$, and
thus has $\bar M +q-1$ zeros at finite $z$.  In addition, the prefactor
multiplying the sum combines with the $M=\bar M$ term to give a zero of
order $q-1-\bar M$ at $\theta=\pi$.  Adding these together, we obtain
the promised result.

    The properties of the $C_\mu^M$ also lead to the
useful identity
$$  \Phi^*_\mu \Phi_\nu - \Phi^*_\nu \Phi_\mu
   = i r^2 \epsilon_{\mu\nu} \Phi^*_\alpha \Phi^\alpha
        \eqno\eq $$
where $\epsilon_{\mu\nu}$ is an antisymmetric tensor whose
whose only nonzero components are
$$  \epsilon_{\theta\phi} = -\epsilon_{\phi\theta} =
      \sin\theta   \, .
    \eqno\eq $$

    We will encounter the quantity $\Phi^*_\mu \Phi^\mu$ in the source
terms for the perturbations of both the electromagnetic field and of
the metric.  Since we will solve these equations by separation of
variables, it is useful to define the expansion
$$  r^2 \Phi_\mu^*(\theta,\phi) \Phi^\mu(\theta,\phi)
       = a^2 \sum_{jm}
    \sigma_{jm} Y_{jm}(\theta,\phi)
     \eqn\Phiexpand $$
where
$$ \sigma_{jm} = {r^2 \over a^2} \int d\phi\, d\theta \sin\theta \,
Y^*_{jm}(\theta,\phi)
   \Phi_\mu^*(\theta,\phi) \Phi^\mu(\theta,\phi)
    \eqn\sigmaintegral $$
and $a$ is as defined in Eq.~\adef.
Ordinary, rather than monopole, spherical harmonics enter here because
we are dealing with a neutral quantity.  Hence, $j$ runs over
integer values although, since $\Phi_\mu$ is a linear
combination of monopole harmonics with angular momentum $\hat J=q-1$,
the $\sigma_{jm}$  vanish for all $j >2(q-1)$.  Note that, as a
consequence of the normalization condition \Cnorm,
$$  \sigma_{00} = {1 \over \sqrt{4\pi}}  \, .
     \eqn\sigzz $$

     The properties of the $J=q-1$ harmonics also simplify
the electromagnetic field equations.  All components
of $j^\nu$ vanish, while
$$ p_{\mu\nu} = {eg\over 2m} \epsilon_{\mu\nu}r^2 f^2 \Phi^*_\alpha
\Phi^\alpha \, .
      \eqno \eq $$

    The various components of Eq.~\Feq\ can be written as
$$ \partial_\mu\left(\sqrt{\bar g}\, \delta F^{\mu t} \right)= 0
       \eqno\eq $$
$$   \partial_\mu\left(\sqrt{\bar g}\, \delta F^{\mu r}\right) = 0
          \eqn\EMra $$
$$  {1\over \sqrt{\bar g} } \partial_\mu\left(\sqrt{\bar g}\,
     \delta F^{\mu a}\right)= -{eg\over 2m}
    \epsilon^{ab} r^2f^2
    \partial_b \left( \Phi_\alpha^* \Phi^\alpha \right)
      \eqn\EMab $$
where we have adopted the
convention that Roman indices from the beginning of the alphabet take
only the values $\theta$ or $\phi$.
(In obtaining the last of these, we have used the fact that
$\sqrt{\bar g}\, \epsilon^{ab}$ is a function only of $r$.)

     Because we are seeking time-independent solutions, the equations for
the electric and magnetic fields decouple.  For the former the source
term vanishes, and so the equations are the same as those encountered in
studying perturbations of the pure Reissner-Nordstr\"om solution, where the
only allowed static perturbation of the electric field is a radial field
corresponding to a variation of the black hole's electric charge.  Since
we are assuming vanishing electric charge, this perturbation must be
excluded, and so $\delta F^{t\mu} =0$.

      The equations for the magnetic field can be solved by separation of
variables.  We first expand $\delta F_{\mu\nu}$ in terms of vector
spherical harmonics:
$$ \eqalign { \delta F_{\theta\phi} &= \sum_{jm}
     F_1^{jm}(r) \,\sin\theta\, Y_{jm}(\theta,\phi) \cr
    \delta F_{ra} &= \sum_{jm} \left[ F_2^{jm}(r) \epsilon_a{}^b \partial_b
         Y_{jm}(\theta,\phi)
         +  F_3^{jm}(r)\, \partial_a Y_{jm}(\theta,\phi) \right]}
    \eqno\eq $$
where it is understood that $F_2^{00} = F_3^{00} =0$.
Substituting this expansion into Eq.~\EMra\ and using the identity
$$ {1 \over \sqrt{\bar g} }\,\partial_\mu  \left( \sqrt{\bar g}\, \bar
     g^{\mu\nu}
  \partial_\nu Y_{jm} \right) = -{j(j+1) \over r^2} Y_{jm}
     \eqno\eq $$
we find that
$$   F_3^{jm} =0  \, .
   \eqno\eq $$

Next, the  $\mu=t$ component of the
Bianchi identity, Eq.~\EMbianchi, leads to
$$  {d F_1^{jm} \over dr}= {j(j+1) \over r^2} F_2^{jm} \, .
    \eqn\Ftwosolve $$
For $j\ge 1$ this can be used to eliminate $F_2^{jm}$, while for $j=0$ it
implies that $ F_1^{00}$ is a constant.  Since a constant $ F_1^{00}$
corresponds to a change in the magnetic charge, we set $ F_1^{00}=0$.

Finally, Eq.~\EMab, together with Eqs.~\Phiexpand\ and
\Ftwosolve, yields
$$  {d\over dr}\left( B {dF_1^{jm} \over dr} \right)
    - { j(j+1) \over r^2} F_1^{jm}
     =- {ega^2\over 2mr^2} j(j+1) {f^2} \sigma_{jm} \, .
   \eqn\Foneeq $$
It will be convenient to write
$$  F_1^{jm}(r)  = {ega^2\over 2} \sigma_{jm} {\cal F}_j(r) \, , \qquad\qquad
j>0\, ,
    \eqn\FonecalF $$
where ${\cal F}_j$ obeys
$$  {d\over dr}\left( B {d{\cal F}_j \over dr} \right)
    - { j(j+1) \over r^2} {\cal F}_j
     =  - {j(j+1)\over mr^2} {f^2 } \, .
    \eqn\calFeq $$

    By multiplying this equation by ${\cal F}_j$ and then integrating over
$r$ from $r_H$ to $\infty$, one can show that for $f=0$ and $j>0$
the only regular solution is the trivial one
${\cal F}_j(r)= 0$.  Hence, in the presence of the source one can solve,
at least formally, for ${\cal F}_j$ by inverting the operator on the left
hand side of Eq.~\calFeq.   To construct the appropriate Green's function
we need the two solutions $g_j^-(r)$ and $g_j^+(r)$ of the homogeneous
equation that are regular at $r=r_H$ and $r=\infty$, respectively.  Using
the fact that $B(r)$ tends to unity at large $r$, one immediately finds
that these two solutions behave asymptotically as $r^{j+1}$ and
$r^{-j}$.  (Explicit forms for these solutions are given in the
appendix.)   If
they are normalized so that $r^{-(j+1)} g_j^-(r)$ and $r^j g_j^+(r)$ both
tend to unity as $r \rightarrow \infty$, then
$$  {\cal F}_j(r) =  - {j(j+1)\over m}
       \int_{r_H}^\infty dr' G_j^{(F)}(r,r') \left[{f(r') \over r'}\right]^2
     \eqn\Greenone $$
where
$$   G_j^{(F)}(r,r')  = -{1\over 2j+1} \left[ \theta(r-r') g_j^+(r) g_j^-(r')
             + \theta(r'-r) g_j^-(r) g_j^+(r') \right]  \, .
   \eqn\Greentwo$$
Note that neither $g_j^+(r)$ nor $g_j^-(r)$ can have any
zeros for $r>r_H$ (i.e., in the region where $B(r)>0$).  This fact,
together with Eqs.~\Greenone\ and \Greentwo, implies that
${\cal F}_j(r)$ is positive everywhere outside the horizon.

    Because $f(r)$ falls exponentially for $r \gg m^{-1}$, the
contribution from the first term in the Green's function dominates at
large distance and so
$$ {\cal F}_j \sim {S_j\over r^j}\, , \qquad r\rightarrow \infty
    \eqn\Fasymp $$
where
$$   S_j =  {j(j+1)\over 2j+1} \int_{r_H}^\infty dr
    {g_j^-(r)f(r)^2 \over mr^2}  \, .
    \eqno\eq $$
As a check that this is indeed the proper behavior, note that this
implies that the $2^j$-pole components of $\delta F_{\theta\phi}$ fall
as $1/r^j$, while the unperturbed monopole component of
$F_{\theta\phi}$ is independent of $r$.
Since $a$ is of
order $\beta/e$,  the magnetic field perturbations
that we have found correspond to magnetic $2^j$-poles with components
equal to the $\sigma_{jm}$ times quantities of order $\beta^2
r_H^j/e \sim \beta^2 /e m^j$.

\chapter{Determination of the $k_M$ and the symmetry of the solution}

      We can now use our results for $V_\mu$ and $\delta F_{\mu\nu}$ to
determine the $k_M$.   The overall scale of these, measured by the
quantity $a$ that was defined in Eq.~\adef, determines the magnitude of
the departure of our solution from the Reissner-Nordstr\"om black hole.
The relative sizes of the various $k_M$ determine the angular
dependence --- i.e., the shape --- of the solution; for studying
these it is convenient to define
$$  n_M = {k_M \over a}
    \eqno\eq $$
that satisfy
$$  \sum_M |n_M|^2 =1  \, .
    \eqno\eq $$

     Substitution of the of the results of the previous section into
Eq.~\Igeneraleq\ gives the quantity
$$  I = -\beta^2 m a^2
 + {\lambda e^2 m a^4 \over 2} p \sum_{jm} \left|\sigma_{jm}\right|^2
    - {e^2g^2m a^4\over 8} \sum_j q_j \sum_m  \left|\sigma_{jm}\right|^2
     \eqn\Ispecial $$
whose minimum determines the $k_M$.  In this expression
$p$ and $q_j$ denote the positive integrals
$$  p= \int dr {f(r)^4 \over m^3 r^2}
    \eqno \eq $$
and
$$   q_j = \int dr {f(r)^2 \over m^2 r^2} {\cal F}_j(r)
    \eqno\eq $$
which are all of order unity.

     Eqs.~\Phidef\ and \Phiexpand\ show that the $\sigma_{jm}$ are
homogeneous polynomials of degree 2 in the $k_M$.  This suggests that
we rewrite Eq.~\Ispecial\ as
$$   I = -\beta^2 m a^2  + {e^2g^2 ma^4 \over 8} I_1(n_M)
   \eqno\eq $$
where
$$  I_1(n_M) = {4\lambda\over g^2} p \sum_{j=0}^{2(q-1)} \Psi_j
    - \sum_{j=1}^{2(q-1)} q_j \Psi_j
        \eqn\Ieq $$
and the rotational scalars
$$ \Psi_j = \sum_{m=-j}^j \left|\sigma_{jm}\right|^2
     \eqn\Psidef $$
are homogeneous polynomials of degree 4 in the $n_M$.
Minimization of $I$ requires
$$   a = {2\beta \over eg} \left[I_1(n_M)\right]^{-1/2}
\, .
     \eqno\eq $$

     The $n_M$ are determined, up to an ambiguity corresponding to the
rotational and global gauge symmetries of the theory, by minimizing $I_1$.
We begin by considering individually several low values of $q$.

\noindent {\bf Case i: $q=1$, $\hat J =0$ }

     The solution with unit magnetic charge is spherically symmetric
(indeed, it is the only case for which spherical symmetry is possible).
There is only a single $n_M$, of unit magnitude, whose phase has no
physical significance.

\noindent {\bf Case ii: $q=3/2$, $\hat J=1/2$}

     There are two
$n_M$ that form a complex $SU(2)\times U(1)$ doublet, where
the former factor refers to spatial rotations and the latter to global
phase rotations of the charged fields.  It is always possible to find
a symmetry transformation that brings such a doublet into the
standard form $n_{1/2} =1$, $n_{-1/2}=0$.   The solution is
axially symmetric in the sense that it is left invariant by a
combination of  a rotation about the
$z$-axis and a global gauge transformation.  In particular, all
gauge-invariant quantities are manifestly axially symmetric.  One finds
that
$$  r^2 \Phi^*_\mu \Phi^\mu  = {a^2 \over 4\pi} (1-\cos\theta)
    \eqno\eq $$
so that the nonzero $\sigma_{jm}$ are
$$  \sigma_{00} = {1 \over \sqrt{4\pi}}\, , \qquad \qquad
             \sigma_{10} =  -{1 \over \sqrt{12\pi}}   \, .
     \eqno\eq $$
The solution has a net magnetic dipole moment that can be
attributed to the asymmetric distribution of the magnetic dipole
density of the charged vector field.

\break

\noindent {\bf Case iii: $q=2$, $\hat J=1$ }

     This case is somewhat less trivial. The three
complex $n_M$ are equivalent to a pair of real vectors $\bf v$ and
$\bf w$ obeying ${\bf v}^2 + {\bf w}^2=1$, with the
correspondence being given by
$$  \eqalign{ n_{\pm 1} &= {1\over \sqrt{2}}
    \left[ \mp(v_x +i w_x) - i(v_y +iw_y) \right] \cr
      n_0 &= v_z +iw_z \, .}
    \eqno\eq $$
Using Eqs.~\sigmaintegral\ and \Psidef\ we find that
$$  \eqalign {\Psi_0 &= {1\over 4\pi} \cr
      \Psi_1 &=  {3\over 4\pi} |{\bf v \times w}|^2  \cr
      \Psi_2 &=  {1\over 20\pi}\left(1 - 3|{\bf v \times w}|^2 \right)}
    \eqno \eq $$
and hence that
$$  I_1 = {1\over 20\pi} \left( {24 \lambda\over g^2} p - q_2 \right)
    + {3 \over 20\pi} \left( {8 \lambda\over g^2} p - 5q_1 +q_2
     \right)   |{\bf v \times w}|^2 \, .
    \eqno\eq $$
The nature of the minimum depends on whether the coefficient of
$|{\bf v \times w}|^2$ is positive or negative.  In the former case, $I_1$
is minimized when $\bf v$ and $\bf w $ are parallel.  By choosing their
direction to be along the $z$-axis and then applying a global phase
rotation, we can bring the solution into the form
$$  n_0=1 \, , \qquad \qquad   n_{\pm 1}=0 \, .
    \eqno\eq$$
It then follows that
$$  r^2 \Phi^*_\mu \Phi^\mu   = {3a^2 \over 8\pi} \sin^2\theta
    \eqno\eq $$
and that
$$  \sigma_{00} = {1\over \sqrt{4\pi}} \, , \qquad \qquad
             \sigma_{20} = - {1\over \sqrt{20\pi}}
     \eqno\eq $$
with all other $\sigma_{jm}$ vanishing.

    If instead the coefficient is negative, then $I_1$ is minimized when
$\bf v$ and $\bf w $ are perpendicular and of equal length.  Any such
solution can be rotated so that $v_x = -w_y=  -1/\sqrt{2}$ with all other
components vanishing.  This gives
$$  n_1=1 \, , \qquad \qquad  n_0= n_{-1}=0
    \eqno\eq$$
and
$$  r^2 \Phi^*_\mu \Phi^\mu   =  {3a^2 \over 16\pi} (1-\cos\theta)^2
\, .
    \eqno\eq $$
The nonzero $\sigma_{jm}$ are
$$    \sigma_{00} = {1\over \sqrt{4\pi}} \, ,   \qquad\qquad
      \sigma_{10} = -{\sqrt{3}\over 4\sqrt{\pi}}\, , \qquad\qquad
      \sigma_{20} = {1 \over 4\sqrt{5\pi} }\, .
     \eqno\eq $$

    Both solutions are axially symmetric; the former is
manifestly invariant under a rotation about the $z$-axis, while the latter
is invariant if the rotation is supplemented by a global gauge
transformation.

\noindent {\bf Case iv: $q=3$, $\hat J=2$}

    For larger $q$, the minima of $I_1$ depend on the actual values of
the integrals $p$ and $q_j$, which we can only determine numerically.
However, if $4\lambda/g^2$ is sufficiently large, the first term in
Eq.~\Ieq\ is dominant and the dependence on the $q_j$ can be ignored to
leading order.  In fact, one only has to minimize
$$ \eqalign {\Sigma & =  \sum_{jm} |\sigma_{jm}|^2 \cr
     &={1\over a^4} \int d\phi\, d\theta \sin\theta \,
         r^4 \left(\Phi^*_\mu \Phi^\mu \right)^2 \, .}
     \eqno\eq $$
The integral in the second line is a sum of integrals of products of
four vector harmonics.  Using the explicit
expressions\footnote{4}{Although these expressions for the vector
harmonics are gauge-dependent, the result for $\Sigma$ is
gauge-independent.}  given in Eq.~\explicit, we obtain
$$ \Sigma  =  \sum_{M_1,M_2,M_3,M_4} A_{M_1,M_2,M_3,M_4} \,
     n_{M_1}n_{M_2}n^*_{M_3}n^*_{M_4}
    \eqno\eq $$
where
$$ A_{M_1,M_2,M_3,M_4} = \delta_{(M_1 + M_2), (M_3+M_4)}
     {[(2q-1)!]^2 (2q + M_1 + M_2 -2)! (2q - M_1 - M_2 -2)! \over
     4\pi(4q-3)! \sqrt{ \prod_{j=1}^4 (q+M_j-1)!(q-M_j-1)! }} \, .
    \eqno\eq $$

    Our problem has now been reduced to the minimization of a
quartic polynomial in $2\hat J +1=5$ complex variables.  Even after
using the rotational and phase freedom to fix some of these, one is
still left with a rather formidable task.   We therefore
used Mathematica to search for minima, finding a solution that can be
rotated into the form
$$   n_0 ={1 \over \sqrt{2} } \, , \qquad \qquad
               n_{\pm 1} =0 \, , \qquad \qquad
               n_{\pm 2} = \pm{ 1\over 2 } \, .
       \eqn\qthreesolution $$
{}From this one finds that the nonzero $\sigma_{jm}$ are
$$ \eqalign{ \sigma_{00} &= {1\over \sqrt{4\pi} }\, , \qquad \qquad
         \sigma_{3,\pm 2} = -{\sqrt{5}\over 4\sqrt{14\pi} } \cr
         \sigma_{40} &= {1\over 24\sqrt{\pi} } \,  , \qquad \qquad
         \sigma_{4,\pm 4} = {\sqrt{70}\over 336\sqrt{\pi} }\, . }
    \eqno\eq $$
This solution has no continuous
rotational symmetry, although it is invariant under the group of
finite rotations that leave the tetrahedron invariant.
This is illustrated by
Fig.~1, where we present a three-dimensional plot of $r^2
\Phi^*_\mu \Phi^\mu $ as a function of angle.
Note that $\Phi_\mu$ vanishes at the
center of each of the faces of the deformed
tetrahedron in this figure, in agreement with our previous
remark that it should have $2(q-1)$ zeros.

     Including the effects of the terms involving the $q_j$  shifts the
location of the minimum of $I_1$. One can verify that
(in contrast with the $q=2$ case) the
$\Psi_j$ contain terms that are linear in the deviations of the $n_M$ from
the values given above.  As a result, the full solution for the $n_M$
changes continuously as  $\lambda/g^2$ is varied.

     Since the minimum found here was obtained by numerical methods,
we do not have an analytic proof that it is in fact the global minimum
(although we are fairly confident that it is.)  However, we can
demonstrate unambiguously that the global minimum is not axially
symmetric.  To do this, we note first that any configuration with all
but one of the $n_M$ equal to zero is invariant under rotations about
the $z$-axis (possibly supplemented by a gauge transformation).  By
evaluating the $\sigma_{jm}$ with $m\ne 0$, it is easy to show that
these are the only configurations with this symmetry.  Explicit
calculations for the five configurations of this form shows that they
all give higher values for $\Sigma$ than does the configuration of
$\qthreesolution$. Hence, the global minimum cannot be achieved by a
configuration that is axially symmetric about the $z$-axis; the
rotational symmetry of the theory then extends this result to an
arbitrary axis of rotation.

\noindent {\bf Larger charges  }

      We have applied the methods used for the $q=3$ case to higher
charges also.  For $q=4$ (i.e., $\hat J=3$), the lowest minimum we
find for $\Sigma$ has
$$     n_{\pm 2} = \pm{1 \over \sqrt{2} }
       \eqno\eq $$
with all other $n_M$ vanishing.  The nonzero $\sigma_{jm}$ are
$$ \eqalign{ &\qquad \quad\qquad \sigma_{00} = {1\over \sqrt{4\pi} } \, , \cr
             \sigma_{40} &= -{7 \over 44\sqrt{\pi} }\, ,\qquad \qquad
    \, \,\,    \sigma_{4,\pm 4} = {\sqrt{70} \over 88\sqrt{\pi} }\, , \cr
 	\sigma_{60} &= -{\sqrt{13} \over 572\sqrt{\pi} }\, ,\qquad\qquad
         \sigma_{6,\pm 4} = -{7\sqrt{13}\over 572\sqrt{14\pi} } \, .}
    \eqno\eq $$
A three-dimensional plot of $r^2 \Phi^*_\mu \Phi^\mu $ for this
solution is shown in Fig.~2.  As suggested by the plot, this solution
is invariant under the discrete rotational symmetries of the cube.
$\Phi_\mu$ has a zero on each face of this roughly cubic shape.

     As we go to higher values of $q$, the solutions develop more
small-scale structure, while at the same time appearing more symmetric
when viewed on a large scale.  (Note that a discrete polyhedral
symmetry such as that exhibited by the $q=3$ and $q=4$ solutions is
impossible for most values of $q$.)
What we see happening is that there is a tendency for the
$2(q-1)$ zeros of $\Phi_\mu$ to be distributed as evenly as possible
over the unit two-sphere.  Lying between these zeros are
maxima of $\Phi^*_\mu \Phi^\mu $.  This behavior can be seen, for
example, in Fig.~3, where we show a solution with $q=12$.

\chapter{Perturbation of the Metric }

    The deviation $h_{\mu\nu}$ of the metric from the
Reissner-Nordstr\"om solution is determined to leading order by the
linearized Einstein equation
$$ \delta G_{\mu\nu}  = -8\pi G  t_{\mu\nu} \, .
     \eqn\Einstein$$
Here $\delta G_{\mu\nu}$ denotes the terms in the Einstein tensor that
are linear in $h_{\mu\nu}$ while $t_{\mu\nu}$ is the leading correction
to the energy-momentum tensor.

     In doing this calculation, we continue to restrict ourselves to
the case where $Gm^2/e^2$ is very small; it was this assumption that
allowed us to decouple the determination of $\delta F_{\mu\nu}$ from
that of $h_{\mu\nu}$.  For our perturbative scheme to be valid, the
horizon radius of the unperturbed solution must be close to $r_{\rm
cr}$, and hence must be of order $m^{-1}$.  When this is the case, the
difference between the Reissner-Nordstr\"om and Schwartzschild metrics
with the same value for $M$ is never greater than order $Gm^2/e^2$
anywhere outside the horizon.  Hence, in solving for $h_{\mu\nu}$ we
can approximate the metric by the corresponding Schwarzschild
metric.  We will consider only
the case $q\ge 1$, $g>0$, so that we
can use the results for the $W$-field and the electromagnetic
perturbations that were obtained in Secs.~4 and 5.

    The first step is to calculate $t_{\mu\nu}$.  The full
energy-momentum tensor can be written as
$$  T_{\mu\nu}  = T_{\mu\nu}^{EM} + T_{\mu\nu}^W
      \eqno\eq $$
where
$$  T_{\mu\nu}^{EM} =  g^{\alpha\beta} F_{\mu\alpha}  F_{\nu\beta}
     -{1\over 4} g_{\mu\nu} g^{\alpha\beta} g^{\lambda\rho}
     F_{\alpha\lambda} F_{\beta\rho}
    \eqn\TEM $$
is the purely electromagnetic part and
$$  \eqalign{ T_{\mu\nu}^W
    &= g^{\alpha\beta}\left[
     W^*_{\mu\alpha}W_{\nu\beta} + W^*_{\nu\alpha}W_{\mu\beta}\right]
    +m^2\left( W^*_\mu W_\nu + W^*_\nu W_\mu\right) \cr
  &\qquad  +{ieg\over 2} g^{\alpha\beta} \left[
     F_{\mu\alpha}  \left( W^*_\nu W_\beta -  W^*_\beta W_\nu \right)
  +  F_{\nu\alpha}  \left( W^*_\mu W_\beta -  W^*_\beta W_\mu \right)
     \right]  \cr  &\qquad
  - g_{\mu\nu} \left[  {1\over 2}  W^*_{\mu\nu}W^{\mu\nu}
        +  m^2 W^*_\mu W^\mu
     +{ieg\over 4} F^{\mu\nu} \left( W^*_\mu W_\nu -W^*_\nu W_\mu\right)
      \right] + O(W^4) \, .}
         \eqn\TW $$
The $O(a^2)$ corrections to $ T_{\mu\nu}^{EM}$ are the
sum of a part linear in $\delta F_{\mu\nu}$ and
a part linear in $h_{\mu\nu}$
that arises from the corrections to the metric in Eq.~\TEM; because the
latter is suppressed by an additional factor of $Gm^2a^2$, we can ignore it
here.  The dominant part of $T_{\mu\nu}^W$, which is also $O(a^2)$, is
obtained by substituting the unperturbed metric and field strength into
Eq.~\TW.  Using Eq.~\RNmagfield\ for the only nonzero component of the
unperturbed electromagetic field strength, together with the results of
Sec.~4, we find that the nonzero components of $t_{\mu\nu} $
can be written as
$$ \eqalign{ t_{tt} &= B\left\{ K_{F1} +    \left[ B(f')^2
   +\left(m^2 -{qg\over 2r^2} \right)f^2 \right] K_W \right\} \cr
    t_{rr} &= - {1\over B}\left\{ K_{F1} + \left[- B(f')^2
   + \left(m^2 -{qg\over 2r^2} \right)f^2 \right] K_W \right\}\cr
    t_{ab}
   &= \bar g_{ab}\left[K_{F1} - {qg\over 2r^2} f^2 K_W \right] \cr
        t_{ra}  &= K_{F2} }
    \eqno\eq $$
where
$$ \eqalign {K_W &= {1\over mr^2} \left( r^2 \Phi^*_\alpha \Phi^\alpha \right)
    =  {a^2\over mr^2} \sum_{jm} \sigma_{jm} Y_{jm}(\theta,\phi) \cr
     K_{F1} &= {q\over e r^4}  \sum_{jm}F_1^{jm}Y_{jm} =
      {qga^2\over 2r^4} \sum_{jm} {\cal F}_j\sigma_{jm}Y_{jm} \cr
    K_{F2}  &= -{q\over e r^4}  \sum_{jm}F_2^{jm}\partial_a Y_{jm}
     = -{qga^2\over 2r^2}
    \sum_{jm}{1\over j(j+1)} {\cal F}'_j\sigma_{jm}\partial_a Y_{jm}
\, .}
    \eqno\eq $$
(In the last two lines we have used Eqs.~\Ftwosolve\ and \FonecalF\ to
relate the perturbations of the field strengths to those of $W_\mu$;
it should be recalled that ${\cal F}_0 =0$.)

    The next step is to expand the components of $h_{\mu\nu}$
in terms of spherical harmonics.  The space-space components of
$h_{\mu\nu}$ can be decomposed into a spin-0 field and a spin-2 field,
the time-time component corresponds to a spin-0 field, and time-space
components can be chosen to vanish because the solution is static.
Thus, there are potentially seven functions of $r$ entering this
expansion for each value of $j$ and $m$.  However, examination of the
parity of $t^W_{\mu\nu}$ and $t^F_{\mu\nu}$ shows that it is
sufficient to consider only those terms corresponding to perturbations
of parity $(-1)^j$. \REF\chandra {S.~Chandrasekhar, {\it The
Mathematical Theory of Black Holes}, (Oxford, 1983).}
(In the terminology of Ref.~\chandra, these are polar perturbations.)
In general, this leaves only five modes for
each value of $j$ and $m$: two, with $l=j$, for the spin-0 fields, and
three, with $l=j-2$, $j$ and $j+ 2$, for the spin-2 field.  Hence,
there are five radial functions, which we define by
$$  \eqalign{  h_{tt} &= B(r)\sum_{jm} H_1^{jm}(r) Y_{jm}(\theta,\phi) \cr
      h_{rr} &= {1\over B(r)}\sum_{jm} H_2^{jm}(r) Y_{jm}(\theta,\phi) \cr
     h_{ab} & =\sum_{jm}\left[ \bar g_{ab} H_3^{jm}(r) Y_{jm}(\theta,\phi) +
        H_4^{jm}(r) \nabla_a\nabla_b Y_{jm}(\theta,\phi) \right]\cr
        h_{ra} &= \sum_{jm} H_5^{jm}(r)\nabla_a Y_{jm}(\theta,\phi)  }
    \eqno\eq $$
where $\nabla_\mu$ denotes the generally covariant derivative with
respect to the unperturbed metric.  For $j=1$
there is no mode with $l=j-2$, and so there should be only four radial
functions.  Indeed, the identity
$$  \nabla_a\nabla_b Y_{1M}(\theta,\phi) = -{\bar g_{ab} \over r^2}
         Y_{1M}(\theta,\phi)
     \eqno\eq $$
shows that the $H_4^{1m}$ is redundant and can be set equal to zero.
Similarly, there should be only three radial functions for $j=0$, and the
fact that $Y_{00}$ is a constant allows us to set $H_4^{00} =H_5^{00} =0$.

     Further simplification can be achieved by utilizing the freedom to
perform coordinate transformations, which change the metric by an amount
$$  \delta_G g_{\mu\nu} = \nabla_\mu e_\nu  + \nabla_\nu e_\mu \, .
     \eqno\eq $$
Writing
$$ \eqalign{  e_t &=0 \cr
              e_r &=  \sum_{jm} f_1^{jm}(r) Y_{jm}(\theta,\phi)\cr
          e_a &=  \sum_{jm} f_2^{jm}(r)\nabla_a Y_{jm}(\theta,\phi) \, ,}
    \eqno\eq  $$
we find that
$$  \eqalign{ \delta_G H_1^{jm} &= - B'f_1^{jm}  \cr
         \delta_G H_2^{jm} &=  2B{f'}_1^{jm} + B'f_1^{jm} \cr
         \delta_G H_3^{jm} &= {2B\over r}f_1^{jm}   \cr
         \delta_G H_4^{jm} &= 2f_2^{jm}  \cr
     \delta_G H_5^{jm} &=  f_1^{jm} + {f'}_2^{jm} -{2\over r}f_2^{jm}
\, .}
   \eqno\eq $$
For $j\ge 2$, we choose $f_2^{jm}$ so that $H_4^{jm}=0$ and then $f_1^{jm}$
so that $H_5^{jm} =0$.  For $j=1$, we choose $f_1^{1m}$ so that $H_1^{1m} =
H_2^{1m}$ and then choose
$f_2^{1m}$ so that $H_5^{1m} =0$.  Finally, for $j=0$ we
choose $f_1^{00}$ (the only coordinate freedom available) to set
$H_3^{00} =0$.

    We next note that $t_{ab}$ is proportional to
$\bar g_{ab}$, even though it could in
principle have also contained terms proportional to  $\nabla_a\nabla_b
Y_{JM}(\theta,\phi)$.  The absence of such terms implies that
$$  \sin^2\theta \, \delta G_{\theta\theta} - \delta G_{\phi\phi}
    ={1\over 2} \sum_{jm} \left(H_2^{jm}-H_1^{jm}\right)
       \left( \sin^2\theta \, \nabla_\theta\nabla_\theta -
          \nabla_\phi\nabla_\phi \right)Y_{jm} =0
    \eqno\eq  $$
from which it follows that
$$ H_1^{jm}  = H_2^{jm}, \quad\quad j\ge 2 \, .
    \eqno\eq  $$

    This leaves only two independent radial functions for each value of $j$
and $m$ and gives us a metric of the form
$$ \eqalign{  g_{tt} &= B(r)\left[ -1 +
    \sum_{j=0}\sum_m H_1^{jm}(r) Y_{jm}(\theta,\phi) \right] \cr
      g_{rr} &= {1\over B(r)} \left[ 1 + H_2^{00}(r)Y_{00} +
   \sum_{j=1}\sum_m H_1^{jm}(r) Y_{jm}(\theta,\phi) \right]\cr
     g_{ab} & =  \bar g_{ab} \left[ 1+ \sum_{j=1}\sum_m
         H_3^{jm}(r) Y_{jm}(\theta,\phi) \right] \, .}
    \eqno\eq $$
All of the $H_a^{jm}$ must vanish as $r \rightarrow \infty$.  At the
horizon, the $tt$ and $ab$ components of the metric, as well as its
determinant, are nonsingular.  We require the same of the perturbed
metric, and hence require that $BH_1^{jm}$, $H_2^{00}-H_1^{00}$, and
$H_3^{jm}$ all be nonsingular at $r=r_H$.  We will see that the field
equations place further restrictions on the behavior near the horizon.

    Differential equations for the various $H_a^{jm}$ are obtained by
expanding both sides of the linearized Einstein equation \Einstein\  in
terms of spherical harmonics. Because of the Bianchi identity,
as well as the symmetry of the problem, many components of the
resulting equations are redundant.  In particular, if we write
$$  \eqalign{ \delta G_{tt}
   &= \sum_{jm} \delta G_{tt}^{jm}(r) Y_{jm}(\theta,\phi) \cr
    \delta  G_{rr}
       &= \sum_{jm} \delta G_{rr}^{jm}(r) Y_{jm}(\theta,\phi) \cr
     \delta G_{ra}
    &= \sum_{jm} \delta G_{ra}^{jm}(r) \nabla_a Y_{jm}(\theta,\phi) }
    \eqno\eq $$
then it is sufficient to calculate
$$ \eqalign{ \delta G_{tt}^{00} &= {B \over r^2}\left[
    (1-rB' -B) (H_1^{00} +H_2^{00})  - H_2^{00}
    - rB (H_2^{00})' \right] \cr
         &= - {B\over r^2} \left(rB H_2^{00}\right)' + O(Gm^2/e^2)  \cr
  \delta G_{rr}^{00} &= {1 \over r^2B} \left[ rB (H_1^{00})' + H_2^{00}
    \right] \cr
  \delta G_{rr}^{jm} &= {1 \over r^2B} \left[ rB (H_1^{jm})'
      - \left(rB + {r^2B' \over 2} \right) (H_3^{jm})'
      + {(j-1)(j+2) \over 2}  (H_3^{jm} - H_1^{jm} ) \right],
     \quad j\ge 1 \cr
   \delta G_{ra}^{jm}
     &= {1\over 2} \left[(H_3^{jm})' - (H_1^{jm})'\right]
       -{B'\over 2B} H_1^{jm},  \qquad j\ge 1 \, .}
    \eqno\eq $$
In the second equality for $\delta G_{tt}^{00}$ we have used the fact that,
with the approximations we are making,
$$   1-rB'(r) -B(r) = O(Gm^2/e^2)
     \eqn\metricidentity $$
everywhere outside the horizon; for a Schwartzschild metric the left hand
side of this equation would vanish identically.

    We start with the $j=0$ modes, for which we can use Eq.~\sigzz.
To leading order, the $tt$ component of
Eq.~\Einstein\ leads to
$$  \left(rBH_2^{00}\right)' = {8\pi G a^2 \over m \sqrt{4\pi}}
      \left[ B(f')^2 + \left(m^2 -{qg\over 2r^2}\right)f^2 \right] \, .
    \eqno\eq $$
A second equation is obtained by
multiplying the $rr$ equation by $B(r)$ and the $tt$ equation by
$1/B(r)$, and then adding these to give
$$    (H_1^{00})' - (H_2^{00})'
        = - {16 \pi Ga^2 \over \sqrt{4 \pi}}  {(f')^2 \over mr} \, .
    \eqno\eq $$
These two equations can be immediately integrated, using the boundary
condition that $H_1^{00}(\infty) = H_2^{00}(\infty)=0$, to give
$$  B(r)H_2^{00}(r) = {2\sqrt{4\pi}G\, \delta M\over r}
   - {2 \sqrt{4\pi}G a^2\over mr}
    \int_r^\infty ds\,  \left[ B(s)(f'(s))^2
    + (m^2 - {qg \over 2s^2} )f^2(s)\right]
    \eqno\eq $$
$$  B(r)H_1^{00}(r)
   = {2\sqrt{4\pi}G\,\delta M\over r}
   - {2\sqrt{4\pi} G a^2\over mr}
    \int_r^\infty ds\,  \left\{ \left[B(s) - {2B(r) r\over s} \right]
     (f'(s))^2  + \left(m^2 -{ qg\over 2s^2} \right) f^2(s) \right\}
    \eqn\Honezerozero $$
where $\delta M $ is an arbitrary constant that may be interpreted as a
shift of the black hole mass. Because $f(r)$ falls exponentially fast
at large distance, so do the integrals on the right hand sides of
these two equations.

    Turning now to the $j \ne 0$ modes, we find that the $ra$
components of Eq.~\Einstein\ give
$$  B(H_3^{jm})' -  B(H_1^{jm})' - B'H_1^{jm} =
   8\pi G a^2\left[{qg\sigma_{jm}\over j(j+1)} \right]
      {B{\cal F}'_j \over r^2 }  , \qquad j\ge 1 \, ,
    \eqn\Graeq $$
while the $rr$ component leads to
$$ \eqalign{& rB(H_1^{jm})' - \left(rB +{r^2B' \over 2} \right)(H_3^{jm})'
       +{(j-1)(j+2) \over 2} \left[H_3^{jm} - H_1^{jm} \right] \cr
   &\qquad    = 8\pi G a^2\sigma_{jm} \left\{ {1\over m}\left[ -B(f')^2
         + \left(m^2 -{qg\over 2r^2}\right)f^2 \right]
         +{qg\over 2r^2} {\cal F}_j \right\}, \qquad j\ge 1 \, .}
   \eqn\Grreq $$
It was noted above that the nonsingularity of $g_{tt}$ required that
$BH_1^{jm}$ be regular at $r=r_H$; this leaves the possibility that
$H_1^{jm}$ might be singular there.  This can be ruled out
by multiplying Eq.~\Graeq\ by $r$ and then adding the result to
Eq.~\Grreq.  The resulting equation can be solved to express
$H_1^{jm}$ in terms of $H_3^{jm}$ and $(H_3^{jm})'$, thus showing that
it is regular at the horizon.  Hence, $(H_1^{jm})'$ is
finite at $r_H$.  Using this fact in Eq.~\Graeq, we find that
$$  H_1^{jm}(r_H) = 0, \qquad j\ge 1 \, .
     \eqn\Honevanishes $$

     For $j=1$, the fact that $H_3^{jm}$ only enters Eqs.~\Graeq\ and
\Grreq\ through its derivative simplifies matters considerably.  By
using Eq.~\Grreq\ to solve for $(H_3^{jm})'$ and then substituting
into Eq.~\Graeq, we obtain a first-order equation involving only
$H_1^{jm}$.  This can be easily integrated, and the result then used
to obtain $H_3^{jm}$.  The two constants of integration are fixed by
the boundary conditions that $H_1^{jm}(r_H) = H_3^{jm}(\infty)=0$.
The result is that
$$\eqalign {  H_1^{1m} &= -{ 8\pi G a^2\sigma_{1m} \over r^2 B(r)}
 \int_{r_H}^r ds \,  {B(s)\over B'(s)} \left\{
   \vphantom{{2\over m} \left[ -B(s) (f'(s))^2
       \left(m^2 - {qg\over s^2}\right)f^2 \right]}
     {qg \over 2s} [ 2B(s) + sB'(s)] {\cal F}'_1(s)
    +{qg\over s^2} {\cal F}_1(s) \right. \cr &\qquad \left.
    + {2\over m} \left[ -B(s) (f'(s))^2
        + \left(m^2 - {qg\over s^2}\right)f^2 \right] \right\} }
     \eqn\Honeone $$
$$ \eqalign {H_3^{1m} &= \int_r^\infty ds \, \left\{ {2\over s} H_1^{1m}(s)
         +{8\pi G a^2\sigma_{1m} \over s B'(s)}
  \left[ \vphantom{{2\over m} \left[ -B(s) (f'(s))^2
       \left(m^2 - {qg\over s^2}\right)f^2 \right]}
{qg\over s^3} \left(sB(s) {\cal F}'_1(s) + {\cal F}_1(s)\right)
            \right.\right.     \cr &\qquad \left.\left.
        +{2\over ms} \left[ -B(s) (f'(s))^2
        + \left(m^2 - {qg\over s^2}\right)f^2 \right] \right] \right\}
\, .}
    \eqn\Honethree $$
To obtain the asymptotic behavior of these expressions, we recall that
$f(r)$ vanishes
exponentially fast at large $r$, while ${\cal F}_1 \sim 1/r$.  This
behavior is just sufficient to guarantee the convergence of the
integral in Eq.~\Honeone, with the result that $H_1^{1m} \sim 1/r^2$.
Inserting this into Eq.~\Honethree\ and again using the asymptotic
behavior of ${\cal F}_j$, we find that $H_3^{1m}$ has the
same asymptotic behavior, and in fact that the difference $ H_1^{1m} -
H_3^{1m} \sim 1/r^3 $.

    In solving for the modes with $j\ge 2$, it is useful to define
$$  T_{jm} = H_1^{jm} - H_3^{jm}
    \eqno\eq $$
and to then rewrite Eq.~\Graeq\ as
$$   H_1^{jm} = - {B \over B'} \left[ T_{jm}' + 8\pi G a^2
\left({qg\sigma_{jm}\over j(j+1)} \right) {{\cal F}'_j\over r^2}
\right] \, .
    \eqn\Honesolve  $$
Substitution of this equation and its derivative into Eq.~\Grreq\
leads to a second order equation involving only $T_{jm}$:
$$  r^2BT''_{jm} +  2(r^2B)'T'_{jm} -(j-1)(j+2)  T_{jm}
   = {16\pi G a^2\sigma_{jm} \over m}
        \left[ -B (f')^2  + m^2f^2 \right]  \, .
    \eqn\Teq $$
(In obtaining this result, both Eqs.~\calFeq\ and \metricidentity\ have
been used, and terms of higher order in $Gm^2/e^2$ dropped.)

    In the absence of the source term there are no nontrivial solutions
for $T_{jm}$ that are regular at both the horizon and spatial infinity.
(This can be readily seen by examining the explicit solutions given in
the appendix.)  Hence, the inhomeogeneous Eq.~\Teq\ can be solved by Green's
function methods similar to those used for Eq.~\calFeq.  Let $h_j^-(r)$
and $h_j^+(r)$ be the solutions of the homogeneous equation that are
regular at $r=r_H$ and $r=\infty$, respectively.  At large $r$ these behave
asymptotically as  $r^{j-1}$ and
$r^{-(j+2)}$.  If they are normalized so that $r^{-(j-1)} h_j^-(r)$ and
$r^{j+2} h_j^+(r)$ both tend to unity as $r \rightarrow \infty$, then the
desired Green's function is
$$   G_j^{(T)}(r,r')  = -{1\over 2j+1}
    \left[ \theta(r-r') h_j^+(r) h_j^-(r')
             + \theta(r'-r) h_j^-(r) h_j^+(r') \right]
   \eqn\TGreentwo$$
and the only regular solution for $T_{jm}$ is
$$  T_{jm}(r) =   {16\pi G a^2 \sigma_{jm}\over m}
       \int_{r_H}^\infty dr' G_j^{(T)}(r,r') (r')^2B(r')
      \left\{-B(r')[f'(r')]^2 +m^2[f(r')]^2 \right\} \, .
     \eqn\TGreenone $$
$H_1^{jm}$ can be obtained immediately from this equation together with
Eqs.~\Greenone\ and \Honesolve.

    At large $r$ the source term in Eq.~\Teq\ is exponentially small,
so the first term in the Green's function dominates
Eq.~\TGreenone; thus
$$ T_{jm} \sim \sigma_{jm}\,{C_j\over r^{j+2} },  \qquad r\rightarrow \infty
   \eqno\eq $$
where
$$   C_j = {16\pi Ga^2\over (2j+1)m} \int_{r_H}^\infty dr\, r^2B
    h_j^- \left[ B(f')^2 -m^2 f^2 \right] \, .
    \eqno\eq  $$
By substituting this result back into Eq.~\Honesolve\ and using
Eq.~\Fasymp\ we obtain the asymptotic behavior
$$ H_1^{jm} \sim {8\pi G a^2\sigma_{jm} \over (2j+1) mr_H}
    {A_j \over r^{j+1} },  \qquad r\rightarrow \infty
   \eqno\eq $$
where
$$   A_j =  \int_{r_H}^\infty dr\,
    \left\{ (2(j+2)r^2B h_j^- \left[B(f')^2 -m^2 f^2 \right]
   + qgj {g_j^- f^2 \over r^2}  \right\}  \, .
     \eqno\eq $$
The resulting large distance behavior of $g_{tt}$
corresponds to that one would obtain from a mass distribution with a
$2^j$-pole moment whose components are equal to the $\sigma_{jm}$ times
quantities of order $\beta^2 r_H^{j-1}/e^2 \sim \beta^2 /e^2 m^{j-1}$.

\chapter{Concluding Remarks}

     In this paper we have exhibited black hole solutions with fields
on the horizon that, contrary
to common expectation, are not spherically symmetric.  As the magnetic
charge increases, the structure of these black holes
becomes more detailed, with higher multipole components appearing in the
long-range electromagnetic and gravitational fields.   At the same time,
the lower multipole moments decrease in magnitude, with the result that the
solutions do begin to approach the expected spherical symmetry, if only
in an averaged sense.

      Although these solutions display some unusual, and
perhaps unexpected, properties, there are rigorous general results
on black holes that they must obey.  We consider here two of these, the
zeroth and second laws of black hole dynamics, that are concerned with the
properties of the black hole horizon.  At constant $t$, the horizon can be
described as the two-dimensional surface
$$ r = r_H + \Delta(\theta,\phi)
     \eqno\eq $$
where $r_H$ is the horizon radius of the unperturbed metric.  To leading
order, the vanishing of $g_{tt}$ on this surface gives
$$  \Delta(\theta,\phi) = {B'(r_H) \over B(r_H)}
      \sum_{jm} H_1^{jm}(r_H)Y_{jm}(\theta,\phi) \, .
    \eqno\eq $$
Eq.~\Honevanishes\ implies that the terms in the sum with $j\ge 1$ all
vanish, so that $\Delta$ is a constant independent of angle.
Using the solution in Eq.~\Honezerozero\ for $H_1^{00}$ (with $\delta
M=0$), we find
$$ \eqalign {\Delta &= - {2 G a^2\over  mr_H B'(r_H)}
   \int^\infty_{r_H} ds\, \left[ B(f')^2
    + \left(m^2-{qg\over s^2} \right) f^2 \right] \cr
    &= 2 a^2\beta^2 G m }
    \eqn\Deltaresult $$
where the second equality follows from the eigenvalue and
normalization Eqs.~\feq\ and \fnorm, as well as Eq.~\metricidentity.
Because the integral in the first line is itself of order $\beta^2$,
our result for $\Delta$ is  proportional to $\beta^4$, rather than
the $\beta^2$ one might have expected.  Since the
the higher order corrections to $W_\mu$ and $\delta F_{\mu\nu}$ could also
shift the horizon by a distance of order $\beta^4$, our result is really
that $\Delta$ vanishes to leading order.

    The zeroth law of black hole dynamics states that the surface gravity
--- which corresponds  quantum mechanically to the black hole temperature
--- is constant over the horizon.  For a stationary black hole the surface
gravity $\kappa$ is given by\ref{R.M.~Wald, {\it General Relativity},
(University of Chicago, Chicago, 1984).}
$$ \kappa^2 = -{1\over 2} (\nabla^\mu \chi^\nu)
     (\nabla_\mu \chi_\nu)
      \eqn\kappadef $$
where the right-hand side is to be evaluated on the horizon and
$\chi_\mu$ is a Killing vector that is orthogonal to the horizon.  If the
metric is actually static, and is written in a manifestly $t$-independent
form with vanishing time-space components $g_{tj}=0$, then
the only nonzero component of this Killing vector is $\chi_t = g_{tt}$
and Eq.~\kappadef\ reduces to
$$ \kappa^2 = -{1\over 4} g^{tt} g^{ij} (\partial_i g_{tt})
      (\partial_j g_{tt})  \, .
     \eqn\kappaexact $$
Expanding this equation to first order in $h_{\mu\nu}$ and taking into
account the fact that the horizon has been shifted by an amount $\Delta$
gives
$$ \eqalign{ \kappa & = {B' \over 2} + {B'' \Delta \over 2} - \partial_r h_{tt}
      + {B' \over 2} \left( B^{-1} h_{tt} - Bh_{rr} \right)  \cr
    &= {B' \over 2} + {B'' \Delta \over 2} - {1\over 2} \sum_{jm}
(BH_1^{jm})'Y_{jm}
          + {B' \over 4} Y_{00} \left(H_1^{00} - H_2^{00} \right)  }
       \eqno\eq $$
where all quantities are to be evaluated at $r=r_H$.  Because of
Eq.~\Honevanishes, all terms in the sum with $j\ge 1$   vanish.  Hence,
$\kappa$ is independent of angle and therefore constant over the horizon,
as required.

     The second law of black hole dynamics is the statement that the area
of a black hole horizon never decreases.   Let us apply this to the case of
an unstable Reissner-Nordstr\"om black hole that is  perturbed and
eventually evolves into a static black hole with hair.  By making the
perturbation sufficiently weak, we can arrange that the mass decrease from
radiation be negligible, so that the final state will be a black hole with
hair that has the same value for $M$ as the original Reissner-Nordstr\"om
solution.  (Indeed, the area law can be used to place an upper limit on the
mass loss from radiation.)  The area of its horizon is
$$  A = \int d\theta \, d\phi \sqrt{g_{\theta\theta}\, g_{\phi\phi} }
    \eqn\areaintegral$$
where the integration is over the surface $r=r_H +\Delta$.  The angular
components $h_{ab}$ of the metric perturbation only have terms involving
involving spherical harmonics with $j\ge 1$.  The contributions linear in
these vanish after the integration over angles, and so for calculating the
leading correction to the area we can replace $g_{ab}$ by the unperturbed
metric $\bar g_{ab}$ and obtain
$$  A = 4\pi (r_H^2 + 2 r_H \Delta) \, .
     \eqno\eq $$
{}From this, together with Eq.~\Deltaresult, we see that the second law is
verified, to leading order.

    The methods we have used to construct our solutions have limited us to
the case where the   horizon radius is close to the critical radius for the
instability of the Reissner-Nordstr\"om solution.  However, there seems to
be every reason to expect that the solutions with smaller horizons will
display similar behavior.  The construction and study of such solutions,
which include new extremal black holes, remains an open problem.

\ack

We are grateful to Kimyeong Lee and Piljin Yi for helpful comments.

\appendix

The Green's functions used to solve Eqs.~\calFeq\ and \Teq\ were
constructed from the solutions of the corresponding homogeneous solutions.
In the former case, the solutions obey
$$  B(r)g''(r) + B'(r)g'(r) - {j(j+1) \over r^2}g(r) = 0 \, .
    \eqn\geq$$
As we explain in Sec.~6, in the approximation to which we are working
$B(r)$ may be replaced by the Schwarzschild metric function $1 - r_H/r$.
It is convenient to define a variable $x =r/r_H$ and rewrite Eq.~\geq\ as
$$  (x^2 -x) {d^2g\over dx^2} + {dg \over dx} - j(j+1) g=0 \, .
    \eqn\geqx $$
This has a polynomial solution of the form
$$  g_j^-(x) =  {(j-1)!(j+1)! \over(2j)!} x^2 P_{j-1}^{(0,2)}(2x-1)
     \eqn\gminus$$
where $P_{j-1}^{(0,2)}(z)$ is a Jacobi polynomial\ref{M.~Abramowitz
and I.A.~Stegun, {\it Handbook of Mathematical Functions},
(Dover, New York, 1965).  Note that the identities needed to obtain
Eqs.~(A.4) and (A.12)
are incorrect in the 1965 printing, but appear correctly in the
1972 printing.}, and the normalization has
been chosen so that $x^{-(j+1)}  g_j^-(x) $ tends to unity as $x
\rightarrow\infty$.  Its value at $x=1$ (i.e., the horizon) is
$$ g_j^-(1) = {(j-1)! (j+1)! \over (2j)!}  \, .
     \eqno\eq $$

    To find the other independent solution, we first note that any two
solutions $g$ and $f$ of Eq.~\geq\ obey
$$  B(r)[g'(r)f(r) - g(r)f'(r) ] = c
  \eqno\eq$$
where $c$ is a constant.   In particular, if we write
$$ g_j^+(x) = g_j^-(x) \ln\left(1 -{1\over x}\right) +k(x) \, ,
   \eqno\eq $$
then $k(x)$ obeys
$$  \left({x-1\over x}\right)
   \left[ g_j^- {dk\over dx} - k{dg_j^- \over dx} \right]
    + {(g_j^-)^2\over  x^2} = c  \, .
   \eqno\eq $$
{}From the fact that $g_j^-(x)$ is equal to $x^2$ times a polynomial of
order $(j-1)$, it follows that this last equation has a solution for $k(x)$
as a polynomial of order $j$.

    Explicit forms for low values of $j$ are
$$   \eqalign{  g_1^-(x) & = x^2       \cr
        g_2^-(x) & = x^2 \left( x - {3\over 4} \right)  \cr
         g_3^-(x) & = x^2 \left( x^2 - {4\over 3} x
       + {2\over 5} \right)}
      \eqno\eq $$
and
$$   \eqalign{  g_1^+(x) & = -3 x^2 \ln\left(1 -{1\over x}\right)
     - 3 x - {3\over 2}  \cr
        g_2^+(x) & = -80 x^2 \left( x - {3\over 4} \right)
               \ln\left(1 -{1\over x}\right)
               - 80 x^2+20 x +{10\over 3}  \cr
         g_3^+(x) & =-1575  x^2 \left( x^2 - {4\over 3} x
       + {2\over 5} \right) \ln\left(1 -{1\over x}\right)
           - 1575 x^3 + {2625\over 2} x^2 - 105 x - {35\over 4}   }
      \eqno\eq $$
where the $g_j^+(x)$ have been normalized so that
$x^j  g_j^+(x) $ tends to unity as $x \rightarrow\infty$.

     With the same approximation for $B$, the homogeneous equation
corresponding to Eq.~\Teq\ becomes
$$  (x^2 -x) {d^2h\over dx^2} +(4x-2){dh \over dx} - (j-1)(j+2) h=0 \, .
    \eqn\geqx $$
This has a polynomial solution of the form
$$  h_j^-(x) =  {(j-1)!(j+1)! \over(2j)!}  P_{j-1}^{(1,1)}(2x-1)
     \eqn\hminus$$
whose value at the horizon is
$$ h_j^-(1) = {j!(j+1)! \over  (2j)! } \, .
    \eqno\eq $$
By methods similar to those used to find $g_j^+(x)$, one finds that the
solution that is regular as $x \rightarrow \infty $ is of the form
$$  h_j^+(x) = h_j^-(x) \ln\left(1 -{1\over x}\right)
  +{\ell(x) \over x(x-1)}
    \eqno\eq $$
where $\ell(x)$ is a $j$th order polynomial.

        Explicit forms for low values of $j$ are
$$   \eqalign{  h_2^-(x) & = x -{1\over 2}       \cr
         h_3^-(x) & = x^2 - x + {1\over 5 }}
      \eqno\eq $$
and
$$   \eqalign{  h_2^+(x) & = 60(2x-1) \ln\left(1 -{1\over x}\right)
      +120 - {10 \over x(x-1)}      \cr
         h_3^+(x) &
     = 35\left\{ 12(5x^2-5x+1) \ln\left(1 -{1\over x}\right)
        +60x -30 +{2x-1 \over x(x-1)} \right\}}
      \eqno\eq $$
where the normalization conventions are analogous to those used for
$g_j^\pm(x)$.

\FIG\fone{A three-dimensional spherical plot of  the quantity
$  r^2 \Phi_\mu^* \Phi^\mu  $ for the $q=3$
solution described in Eq.~(5.24).
If we denote the spherical coordinates
of a point as $(R, \theta, \phi)$, then this plot shows the surface
$R(\theta,\phi) =  r^2 \Phi_\mu^* (\theta,\phi)\Phi^\mu (\theta,\phi)$.
Note that while $\theta$ and $\phi$ represent the corresponding spatial
coordinates, $R$ is unrelated to any physical spacetime coordinate.
Note the tetrahedral symmetry of the surface.}

\FIG\ftwo{A spherical plot, similar to that shown in Fig.~1, for the $q=4$
solution of Eq.~(5.26).  The cubical symmetry of the solution is apparent.}

\FIG\fthreea{(a)\quad A spherical plot, similar to those in
Figs.~1 and 2, for a solution with $q=12$. There is no apparent
symmetry. \hfil\break
(b)\quad Another presentation of the same solution.
The value of the function  $ r^2 \Phi_\mu^* \Phi^\mu  $ on the
unit hemisphere ($0 \le \phi \le \pi$)
is represented by one of 16 gray levels, with black
being the minimum (zero) and white being the maximum.  One can see the
fairly even distribution of the zeros.}

\refout
\figout

\end